\documentclass{main}
\usepackage{subfigure}
\usepackage{dcolumn}
\usepackage{bm}
\usepackage{epsfig}
\usepackage{amsfonts}
\usepackage{amssymb,amscd}
\usepackage{soul}
\usepackage{color}
\def\be{\begin{equation}}
\def\ee{\end{equation}}
\def\bea{\begin{eqnarray}}
\def\eea{\end{eqnarray}}

\newcommand{\ra}{\rangle}

\newcommand{\Photo}{}

\begin{document}

\title{\Large Novel aspects of particle production \\
in ultra-peripheral collisions}

%

\author{F.C. Sobrinho$^{1}$, L.M. Abreu$^{1,2}$, C.A. Bertulani$^{3,4}$,
I. Danhoni$^4$, F.S. Navarra$^{1}$\\
\vskip0.5cm
$^1$Instituto de F\'{\i}sica, Universidade de S\~{a}o Paulo,
Rua do Mat\~ao 1371, Cidade Universit\'aria, -  05508-090,\\ 
S\~{a}o Paulo, SP, Brazil\\
$^2$Instituto de F\'{\i}sica, Universidade Federal da Bahia,
Campus Ondina - 40170-115,  \\ Salvador, Bahia, Brazil\\
$^3$Department of Physics and Astronomy, Texas A\&M University-Commerce,\\
Commerce, Texas 75429, USA\\
$^4$Institut f\"ur Kernphysik,
Technische Universit\"at Darmstadt, 64289 Darmstadt, Germany\\
}

\maketitle

\vskip1.0cm

\abstracts{One of the hot topics in hadron physics is the study of the new 
exotic charmonium states and the determination of their internal struture. 
Another important topic is the search for effects of the magnetic field 
created in high energy nuclear collisions. In this note 
we show that we can use ultra-peripheral collisions to address both topics.  
We compute the cross section for the production of the $D^+ D^-$ molecular 
bound state in $\gamma-\gamma$ collisions. We also show how the magnetic 
field of the projectile can induce pion production in the target. 
Both processes have sizeable cross  
sections and their measurement would be very useful in the study of the 
topics mentioned above.}


\section{Introduction}

In ultra-peripheral collisions (UPCs) target and projectile do not overlap 
and stay intact. 
As a consequence only few particles are produced, the                   
background is reduced and we can study more carefully specific particle 
production processes, such as those addressed here. In UPCs the elementary 
processes which contribute to particle production are photon-photon, 
photon-Pomeron and Pomeron-Pomeron fusion. They are a good environment to 
search for particles which are more difficult to identify in central 
collisions \cite{upc}. 

In this work we discuss two processes of particle production, which may be 
studied in UPCs: production of $D^+ D^-$ meson   
molecules and production of forward pions.  In the first we can gain some 
insight on the nature of these exotic charmonium states and in the second 
we can measure the magnetic field produced by relativistic heavy ions. 

\section{Production of charm meson molecules}

One important research topic in modern hadron physics is the
study of the exotic charmonium states \cite{exo}. These new mesonic
states are not conventional $c \bar{c}$ configurations and their minimum
quark content is  $c \bar{c} q \bar{q}$. The main question
in the field is: are these multiquark states compact tetraquarks or are they 
large and loosely bound meson molecules? Perhaps the largest fraction of the 
community tends to believe that they are molecules. One of the frequently 
invoked arguments is
that the masses of almost all these states are very close to thresholds, i.e. 
to the sum of the masses of two well known meson states \cite{exo,pi,pp1}. 
A genuine tetraquark state could in principle have any mass, including masses 
far from thresholds. Besides, some problems have been detected in the 
calculation of tetraquark masses with QCD sum rules \cite{lucha,rica}. 
Nevertheless, so far there is no conclusive answer. 

The production of hadron molecules has been discussed in the context of B
decays \cite{pi}, in $e^+ e^-$ collisions,  in proton-proton
\cite{pp1,pp2,pp3}, in proton-nucleus, in central nucleus-nucleus
collisions \cite{aa} and also in UPCs \cite{br}. 
In this section we focus on the $D^+ D^-$ molecule   
production in UPCs, but  
the method employed here is applicable to all molecular states. 

The $D^+ D^-$ pair is produced from two photons.
This process can be described by  well known
hadronic effective Lagrangians, from which we obtain the pair production
amplitude. This amplitude is subsequently projected onto the amplitude for
bound state formation. If the properties of the bound state are known, the
only unknown in this formalism is the form factor, which must be attached to
the vertices to account for the finite size of the hadrons.

\begin{figure}                                                  
\begin{tabular}{cc}                                                           
\includegraphics[width=.45\linewidth]{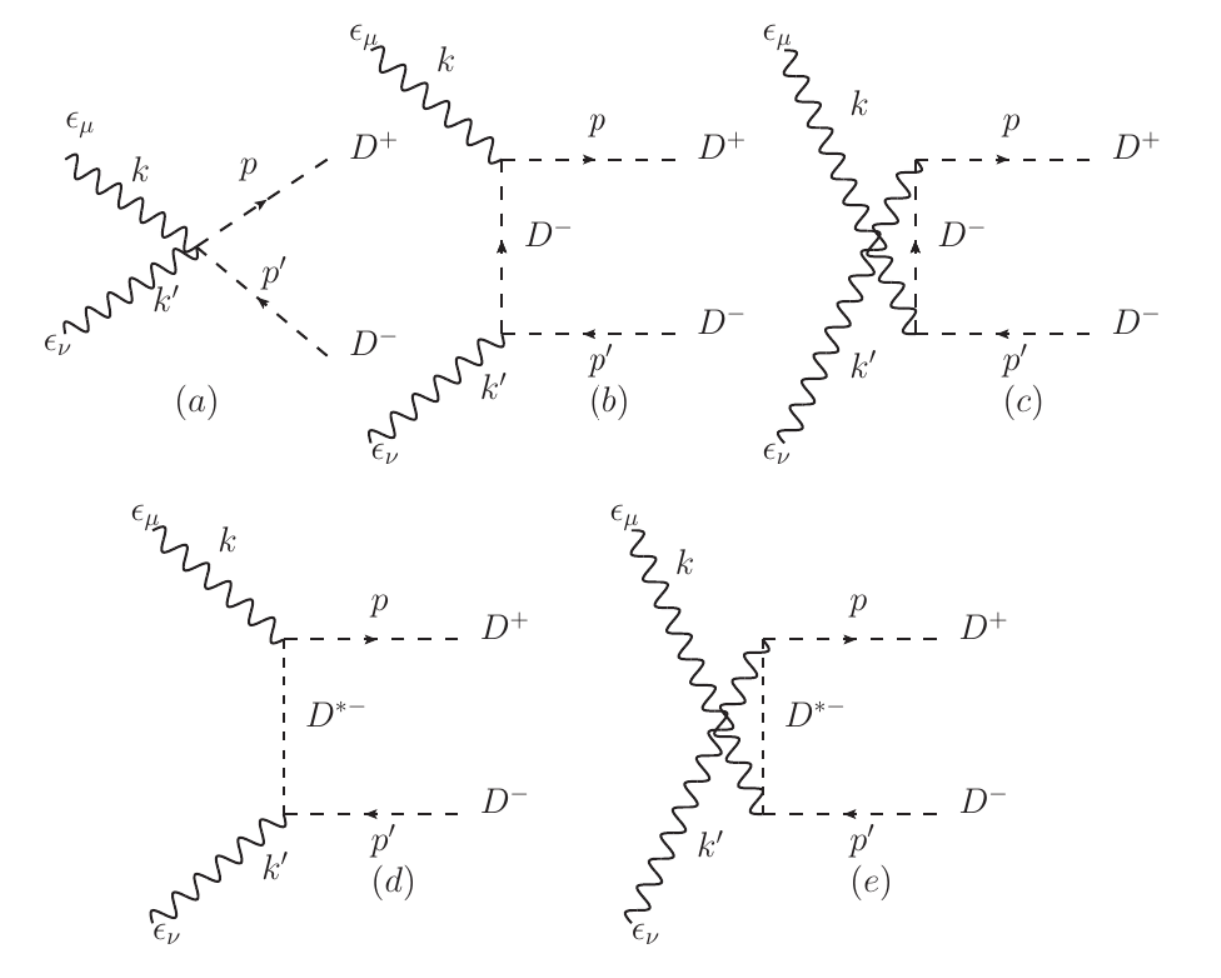}&                   
\includegraphics[width=.45\linewidth]{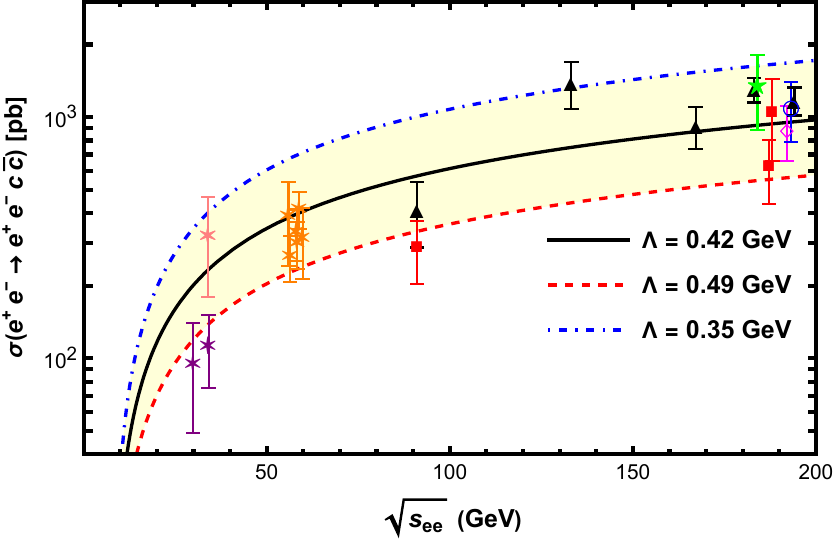} \\               
  (a) & (b)                                               
\end{tabular}                                  
 \caption{a) Feynman diagrams for the process $\gamma\gamma\rightarrow D^+D^-$.
b) Energy dependence of the $e^+ e^- \to e^+ e^- c \bar{c}$ cross section. 
Data come from several collaborations from LEP and were taken from   
arXiv:hep-ex/0010060 [hep-ex]. Lines represent the results obtained with 
Eq.(\ref{sigmafp2}) adapted to $e^+ e^- \to e^+ e^-  D^+D^-$.}    
\label{fig1}                                                                   
\end{figure}

We will study the process
$\gamma \gamma \to D^+ D ^-$ with the Lagrangian densities \cite{lag}
\begin{equation}
    \mathcal{L} = (D_\mu\phi)^*(D^\mu\phi) - m_D^2\phi^*\phi -
\frac{1}{4}F_{\mu\nu}F^{\mu\nu} \, ,
\end{equation}
and
\begin{equation}
\mathcal{L} = -ig_{\gamma D^{+}D^{*-}} F_{\mu\nu}\epsilon^{\mu\nu\alpha\beta}
(D^{*-}_{\alpha}\partial_{\beta}D^{+} - \partial_{\beta} D^{*-}_{\alpha} D^{+} 
+D^{-}\partial_{\beta}D^{*+}_{\alpha} - \partial_{\beta}D^{-} D^{*+}_{\alpha})
\, ,
\end{equation}
where
\begin{equation}
D_\mu\phi = \partial_\mu\phi + ieA_\mu\phi \,\, ,    \hskip1cm
F_{\mu\nu} = \partial_\mu A_\nu - \partial_\nu A_\mu \nonumber \, ,
\end{equation}
and $\phi$, $D^*$ and $A_{\mu}$ represent the pseudoscalar charm meson 
(with mass $m_D$), the vector charm meson (with mass $m_D^*$) and the
photon field, respectively. The Feynman rules can be derived from the
interaction terms and they yield the Feynman diagrams for the process
$\gamma \gamma \to D^+ D^-$ shown in Fig.~\ref{fig1}a.  In the figure
we also show the quadrimomenta of the incoming photons
$k^{\mu} = (E_{p},0,0,{\bf k})$, ${k'}^{\mu} = (E_{k'},0,0,{\bf k'})$
and of the outgoing mesons $p^{\mu} = (E_p,0,0,{\bf p})$,
${p'}^{\mu} = (E_{p'},0,0,{\bf p'})$. The scattering amplitude can be derived 
from the Feynman rules. 

As usual, we  include form factors, $F(q)$, in the vertices of the amplitudes.
We shall follow \cite{kk} and use the monopole form factor given by
\begin{equation}
F(q^2) = \frac{\Lambda^2 - m^2_{D^{(*)}} }{\Lambda^2 - q^2} \, ,
\end{equation}
where $q$ is the 4-momentum of the exchanged meson and $\Lambda$ is a cut-off
parameter.  This choice has the advantage of yielding
automatically $F(m^2_D) = 1$ and $F(m^2_{D^*}) = 1$ 
when the exchanged meson is on-shell. The above form is arbitrary but there is
hope to improve this ingredient of the calculation using QCD sum rules to
calculate the form factor, as done in \cite{ff}, thereby reducing the     
uncertainties. Taking the square of the amplitude and the average over the 
photon polarizations it is straigthforward to calculate the cross section:
\begin{equation}
\label{crossfp}
\sigma = \frac{1}{64\pi^2}\frac{1}{\hat{s}}\sqrt{1 - \frac{4m_D^2}{\hat{s}}}
\int |M(\gamma \gamma\rightarrow D^+D^-)|^2d\Omega\, .
 \end{equation}
where $ \hat{s}=(k + k')^2$. 
We emphasize that the only unknown in our calculation is the cut-off
parameter $\Lambda$. In what follows, we will determine it  fitting our cross
section to the LEP data on the process $e^+ e^- \to e^+ e^- c \bar{c}$. 

From the $D^+ D^-$ pair we can  construct a bound state (denoted $B$).
As in \cite{pp1}, we impose phase space constraints on
the mesons, forcing them to be ``close together''. Here we do this through the
prescription discussed in \cite{pes}. The bound state $|B \rangle$ is
defined as
\begin{equation}
\frac{|B\rangle}{\sqrt{2E_B}}\equiv \int
\frac{d^3q}{(2\pi)^3}\tilde{\psi^*}({\bf q})
\frac{1}{\sqrt{2E_q}}\frac{1}{\sqrt{2E_{-q}}}|{{\bf q},-{\bf q}\rangle },
\label{bound}
\end{equation}
where $E_B$ is the bound state energy, ${\bf q}$ is the relative three momentum
between  $D^+$ and $D^-$ in the state $B$, $E_{\pm q}$ are the energies of
$D^+$ and $D^-$ and $\tilde{\psi}({\bf q})$ is the bound state wave function   
in momentum space. From Eq. (\ref{bound}), we can write the following relation 
between the amplitudes:
\begin{equation}
\frac{M(\gamma\gamma\rightarrow B)}{\sqrt{2E_B}}= \int \frac{d^3q}{(2\pi)^3}
\tilde{\psi}^*({\bf q})\frac{1}{\sqrt{2E_{D^+}}}\frac{1}{\sqrt{2E_{D^-}}}
M(\gamma\gamma\rightarrow D^+D^-),
\end{equation}
We assume that the  ${\bf p} \simeq {\bf p'}$ and hence
$E_{D^+} \simeq E_{D^-} = E_D$. Consequently, the relative momentum
${\bf q} = {\bf p} - {\bf p'} $ is close to zero. Therefore the energy 
$E_D$ and the amplitude $M(\gamma\gamma\rightarrow D^+D^-)$
depend only weakly on  ${\bf q}$ and can be taken out of the integral.
Moreover,
since the binding energy is small we have $E_B \simeq 2E_D$ and hence
\begin{equation}
M(\gamma\gamma\rightarrow B) 
= \psi^*(0)\sqrt{\frac{2}{E_B}}M(\gamma\gamma\rightarrow D^+D^-) \, .
\label{project}
\end{equation} 
With the amplitude above we calculate the cross section for bound state
production:
\begin{equation}
d\sigma = \frac{1}{H}\frac{d^3p_B}{(2\pi)^3}\frac{1}{2E_B}(2\pi)^4
\delta^{(4)}(k+k'-p_B)|M(\gamma\gamma\rightarrow B)|^2,
\label{sigbs}
\end{equation}
where $p_B$ is the momentum  of the produced bound state and $H$ is 
the flux factor.  In the center of mass  frame of the $AA \to AAB$ 
collision, we have
\begin{equation}
k = (\omega_1,0,0,\omega_1) \, , \hskip1.0cm
k' = (\omega_2,0,0,-\omega_2) \, ,  \hskip1.0cm
p_B \equiv p+p' = (E_B,0,0,\omega_1-\omega_2) \, ,
\end{equation}
where $E_B = \sqrt{(\omega_1-\omega_2)^2 + m_B^2}$ and $\omega_1$ and
$\omega_2$ are the energies of the colliding photons.
The flux factor is then given by $H = 8 \omega_1\omega_2$. 
The integrated cross section reads:
\begin{equation}
\sigma(\omega_1,\omega_2) = \frac{2\pi}{2(4\omega_1\omega_2)}\int
\frac{d^3p_B}{2E_B}\delta(E_{CM} - E_B)\delta^{(3)}({\bf k}+{\bf k'}-
{\bf p}_B)\left[\frac{2}{E_B}|\psi(0)|^2|M(\gamma\gamma\rightarrow D^+D^-)
|^2\right]
\label{sigmabound2}
\end{equation}
where  $E^2_{CM} = 4 \omega_1 \omega_2$. 
To complete the calculation we need the wave function of the bound state. 
In  \cite{go} a similar particle made  of open charm mesons was
studied with the Bethe-Salpeter equation and an expression for the wave
function was derived. Here we will just quote the final expression needed to
calculate $\psi(0)$, which is given by:
\be
\psi(0) = \frac{-8 \mu \pi g}{(2\pi)^{3/2}} 
\left(\Lambda_0 - \sqrt{2\mu E_{b}} \arctan\left(\frac{\Lambda_0}
{\sqrt{2\mu E_{b}}}\right)\right) \, ,
\hskip1cm
g^2= \frac{\sqrt{2\mu E_{b}}}{8\pi\mu^2
(\arctan(\frac{\Lambda_0}{\sqrt{2\mu E_{b}}}) - \frac{\sqrt{2\mu E_{b}} 
\Lambda_0}{2\mu E_{b} + \Lambda_0^2})}.
\label{gs} 
\ee
In the above expressions $\mu$ is the reduced mass ($\mu = m_D /2$),
$\Lambda_0$ is a cut-off parameter and $E_{b}$ is the binding energy.
We shall follow \cite{mb} and assume that $\Lambda_0 = 1$ GeV. From \cite{mb} 
we see that one can compute the (dynamically generated) mass of a bound state 
and then determine its binding energy. Knowing $\mu$, $E_b$
and fixing $\Lambda_0$, we can use (\ref{gs}) to calculate $\psi(0)$.
In what follows our reference value will be obtained using  $m_D =1870$ MeV
and the mass of the bound state equal to  $M_B = 3723$ MeV, as  found in
\cite{mb}. With these numbers we get $E_{b}=17$ MeV and 
$|\psi(0)|^2 = 0.008$ GeV$^3$.

The equivalent photon approximation is well known and it is described in
several papers \cite{epa}. In general, when the photon source is a nucleus
one has to use form factors and the calculation becomes somewhat complicated.
Here we will follow \cite{vy} and define an UPC in momentum space.
The momentum distribution of the equivalent photons created by a source with 
charge $Ze$ is \cite{vy}:
\begin{equation}
n(\vec{q})  = \frac{Z^2 \alpha}{\pi^2 \omega}
\frac{(\vec{q})^2}{\left( (\vec{q})^2
+ (\omega/\gamma)^2\right)^2} 
\label{defn}
\end{equation}
where $\vec{q}$ is the photon transverse momentum, $\omega$  
its energy and $\gamma$ is given by $\gamma = \sqrt{s}/2m_p$, 
where $m_p$ is the proton mass.
In order to  obtain the energy spectrum, one has to integrate this expression
over the transverse momentum up to some value $\hat{q}$. The value of $\hat{q}$
is given by $\hat{q} = \hbar c/ 2R$, where $R$ is the radius of the projetile.
For Pb, $R \approx 7$ fm and hence $\hat{q} \approx 0.014$ GeV. After the
integration over the photon transverse momentum the  photon energy
distribution  is given by:
\begin{equation}
n(\omega) = \frac{2Z^2\alpha}{\pi}\ln\left(\frac{\hat{q}\gamma}{\omega}
\right)\frac{1}{\omega},
\label{nomega}
\end{equation} 
Because of the approximations \cite{vy} the above distribution is valid when the
condition  $\omega \ll \hat{q}\gamma$
is fullfiled. Using Eq. (\ref{nomega}) we can compute  the cross sections of
free pair production, $\sigma_P$, and of bound state production, $\sigma_B$.
They are given by: 

\begin{equation}
\sigma_P(A \, A \rightarrow A \, A \,  D^+D^-)
= \int\limits_{m_D^2/\hat{q}\gamma}^
{\hat{q}\gamma}  d\omega_1 \int\limits_{m_D^2/\omega_1}^{\hat{q}\gamma}
d\omega_2 \, \sigma_{P}(\omega_1,\omega_2) \, n(\omega_1) \, n(\omega_2),
\label{sigmafp2}
\end{equation}
\begin{equation}
\sigma_B (A \, A \rightarrow A \, A \,  B) 
= \int\limits_{m_D^2/\hat{q}\gamma}^
{\hat{q}\gamma}d\omega_1\int\limits_{m_D^2/\omega_1}^{\hat{q}\gamma}
d\omega_2 \, \sigma_{B}(\omega_1,\omega_2) \, n(\omega_1) \, n(\omega_2),
\label{sigmabs2}
\end{equation}
where $\sigma_{P}(\omega_1,\omega_2)$ and $\sigma_{B}(\omega_1,\omega_2)$ are
given by Eqs. (\ref{crossfp}) (with $\hat{s} = 4 \omega_1 \omega_2$)
and (\ref{sigmabound2}) respectively.
\begin{figure}
\begin{tabular}{cc}
\includegraphics[width=.45\linewidth]{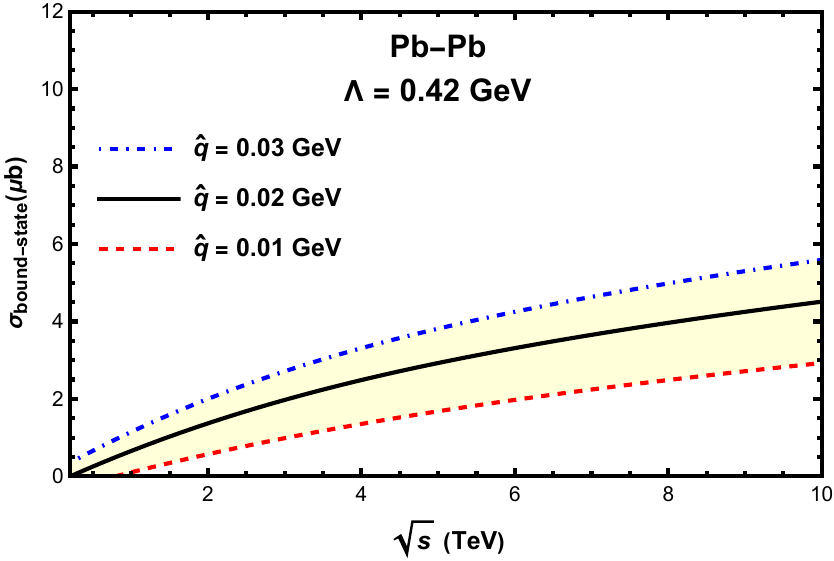}&
\includegraphics[width=.45\linewidth]{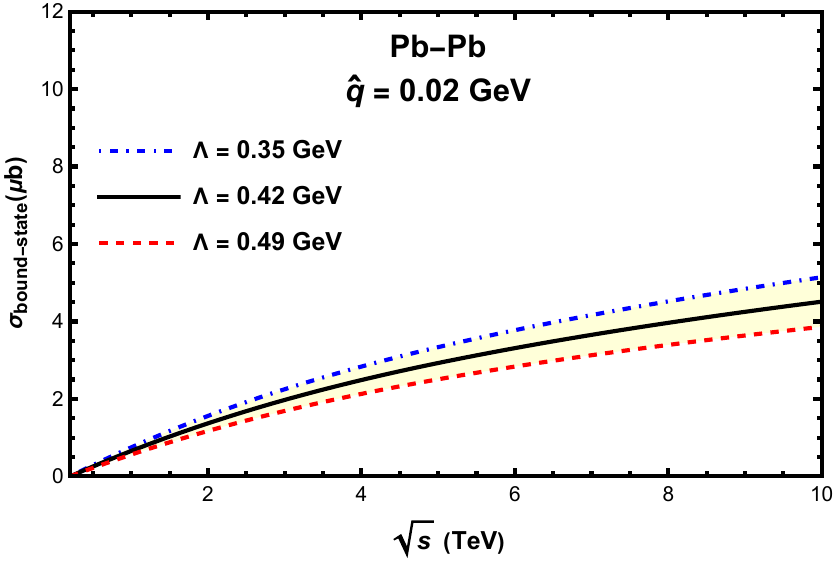} \\ 
(a) & (b) 
\end{tabular}
\caption{Cross sections for $D^+ D^-$ bound state production as a function
of the energy $\sqrt{s}$. a) Dependence on $\hat{q}$ for fixed $\Lambda$.
b) Dependence on $\Lambda$ for fixed $\hat{q}$.
}
\label{fig2}
\end{figure}


In Fig.~\ref{fig1}b we show the cross
sections for free pair production and compare it to the existing experimental
data from LEP \cite{lep}. In fact, the LEP data are for
$e^+ \, e^- \to e^+ \, e^- \, c \, \bar{c}$, i.e., the measured final states are
$D^+ D^-$ and $D^0 \bar{D}^0$.  We assume that these two final states have the
same cross section and, in order to compare with the data, we multiply our
cross section $ \sigma( e^+ \, e^- \to e^+ \, e^- \, D^+ \, D^-) $ by a factor
two. In order to fit these data we
will adapt expression (\ref{sigmafp2}) to  electron-positron collisions. The
$\gamma \gamma \to D^+ D^-$ cross section is the same but the  photon flux
from the electron (and also from the positron) and the integration limits are
different \cite{upc,epa,vy}. Comparing our formula with
these data, we determine the only parameter in the calculation, which is the
cut-off $\Lambda$. In the figure, the curves are obtained  substituting
Eqs. (\ref{crossfp}) and (\ref{nomega}) into (\ref{sigmafp2}). In the latter
$\hat{q} = m_e$.  The band is defined by the choice of two limiting values of
$\Lambda$. In what follows we will use these values to estimate the 
uncertainty of our results. 

In Fig.~\ref{fig2} shows the cross section for bound state production cross 
section and  its dependence on $\hat{q}$ (Fig.~\ref{fig2}a) and 
on $\Lambda$  (Fig.~\ref{fig2}b). It is encouraging to see that 
at $\sqrt{s_{NN}} \approx 5.02$ TeV we have:
\begin{equation}
\sigma (Pb \, Pb \rightarrow Pb \, Pb \, B) =
3.0^{+0.8}_{-1.2} \,\, \mu b
\label{sigbfinal}
\end{equation} 
This number should be compared with results found in \cite{br} and in
\cite{fa}. In those papers, the production cross section of scalar
states $X(3940)$ and $X(3915)$ in $Pb-Pb$  UPCs  at
$\sqrt{s_{NN}} = 5.02$ TeV were calculated and the results were in the range
\begin{equation}
5 \leq \sigma (Pb \, Pb \rightarrow Pb \, Pb \, R) \leq 11 \,\, \mu b
\label{sigx}
\end{equation}
where $R$ stands for $X(3940)$ or $X(3915)$. In  both papers the X states were
treated as meson molecules, as in the present work. It is reassuring to see 
that, in spite of the differences, the obtained cross sections are not so 
different. The cross section (\ref{sigbfinal}) should also be compared 
with the results obtained in \cite{esp} for the production of the same state 
treated as a tetraquark. Interestingly in that work the authors find 
$\sigma (Pb \, Pb \rightarrow Pb \, Pb \, B) = 0.18 \,\, \mu b$, more than one 
order of magnitude smaller than (\ref{sigbfinal}). The existence of this 
significant difference is auspicious for our scientific goal, namely, to use 
UPCs to discriminate between hadron molecules and tetraquarks. 
For more information about the material presented in this 
section we refer the reader to the article \cite{nos24}.

\section{Production of very forward pions}

Some time ago, 
several calculations \cite{skokov,voro} have shown that in high energy nuclear 
collisions a very strong  magnetic field is produced. Since then, the effects 
of these fields have been looked for in different physical processes. Perhaps
the most famous one is the chiral magnetic effect \cite{cme}. Another effect 
was investigated in \cite{nos20}.  In that work it was argued that in an
ultra-peripheral collision magnetic excitation (ME) can lead to pion 
production.
In a ME the projectile creates a $\vec{B}$ field which causes a ``splin-flip'' 
in a  nucleon in the target, resulting in the process $N \to \Delta$. 
After the excitation, the 
$\Delta$ decays into pions: $\Delta \to N + \pi$.  The pions produced in this 
way have extremely large rapidities. This is in contrast to all other particle 
production processes in UPC, in which the produced particles have small 
rapidities. Hence the appearance of very forward pions would be               
a confirmation of ME and would be a clear manifestation of the magnetic field.
In \cite{nos20} and \cite{nos21}  this process was treated in two different 
ways.
\begin{figure}[h]
\centering
\vskip0.5cm
\includegraphics[width=.70\linewidth]{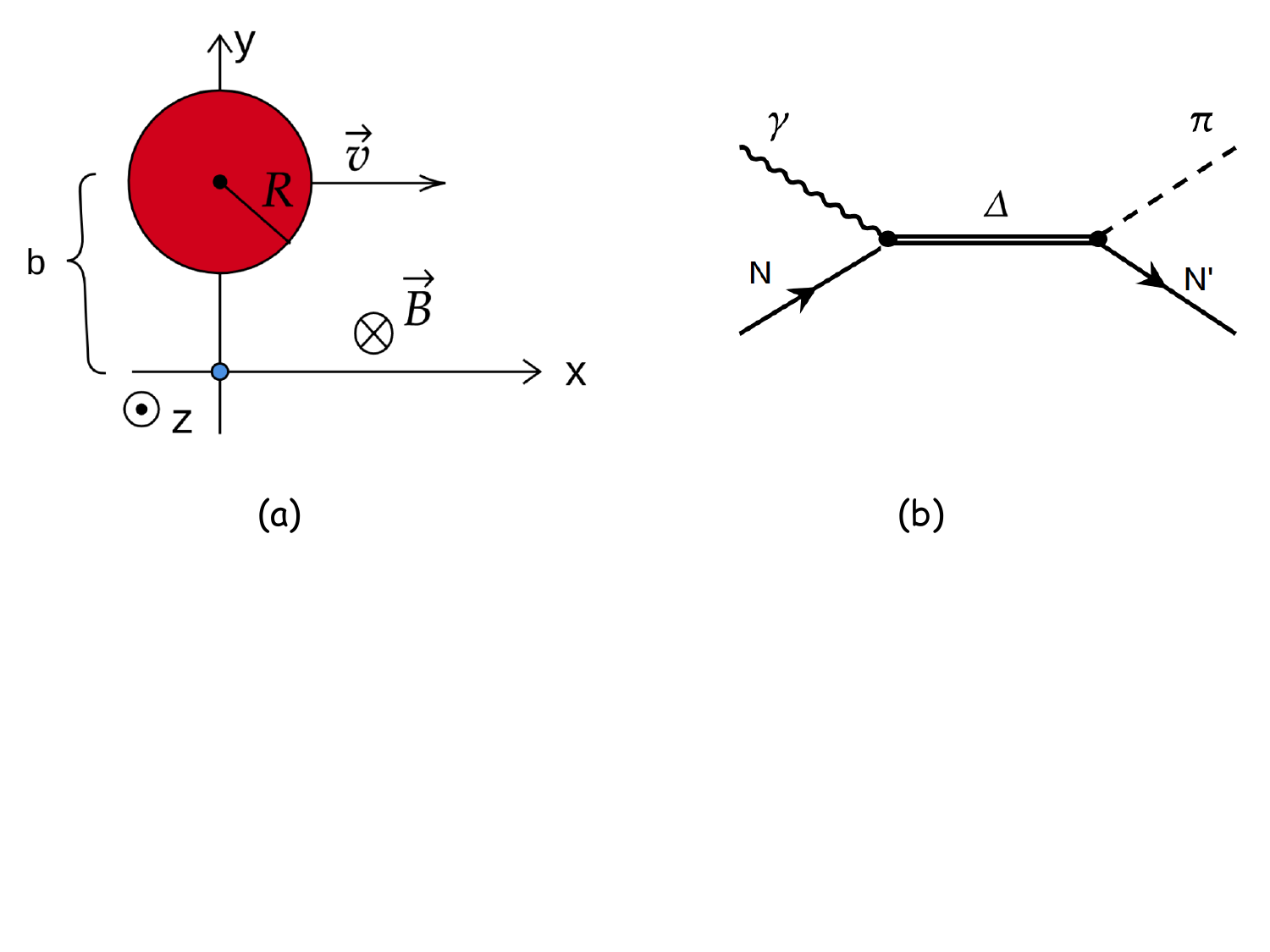}
\vskip-3.5cm
\caption{a) Classical magnetic transition: a moving projectile creates a
magnetic field $\vec{B}$ which acts on the target at rest (at the origin of 
coordinates) flipping its spin.
 b) Quantum version of the  same transition.}
\label{fig3}
\end{figure}
In what folows we will briefly review the two calculations and compare them.

Let us start considering the $Pb - p$ collision depicted in 
Fig.~\ref{fig3}a, where the proton is at rest. The incoming nucleus creates a
magnetic field which converts the proton  into a $\Delta$. As an example, 
let us focus on  the
transition $|p \uparrow \rangle  \to | \Delta^+ \uparrow \rangle$.
The corresponding amplitude  reads \cite{nos20}:
\begin{equation}
a_{fi} = -i \int_{-\infty}^{\infty} e^{iE_{fi}t'}
\langle \Delta^+ \uparrow| H (t') |p \uparrow \rangle \, dt'
\label{amp}
\end{equation}
where $E_{fi}=(m_\Delta^2-m_p^2)/2m_p$,  
$m_{\Delta}$ is the $\Delta$ mass and $m_p$ is the proton mass. 
The Hamiltonian reads:
\begin{equation}
H (t) = - \vec \mu . \vec B (t)  \,\,\,\,\,\,\,\,\,\,\,\,\,\,
  \mbox{with}
  \,\,\,\,\,\,\,\,\,\,\,\,\,\,
\vec \mu = \sum_{i=u,d}\vec \mu_i = \sum_{i=u,d} \frac{q_i}{m_i}\vec S_i
\label{hint}
\end{equation}
where $q_i$ and $m_i$ are the charge and the mass of the constituent quark of 
type $i$ and
$\vec{S}_i$ is the spin operator acting on the spin state of  this quark.

The system of coordinates is shown in Fig. \ref{fig3}a.   
The $\vec{B}$ field is given by \cite{nos20}:
\begin{equation}
   B_z (t) = \frac{1}{4\pi}\frac{qv\gamma (b-y)}{((\gamma(x-vt))^2
        +(y-b)^2+z^2)^{3/2}}
\label{field}
\end{equation} 
where  $v \simeq 1$ and  $q =Z e$. The required spin wave functions are:
\begin{equation}
|p \uparrow \ra
=\frac{1}{3\sqrt{2}}[udu(\downarrow\uparrow\uparrow+
  \uparrow\uparrow\downarrow
    -2\uparrow\downarrow\uparrow)+duu(\uparrow\downarrow\uparrow
    +\uparrow\uparrow\downarrow
    -2\downarrow\uparrow\uparrow)+uud(\uparrow\downarrow\uparrow
    +\downarrow\uparrow\uparrow
-2\uparrow\uparrow\downarrow)]
\end{equation}
\begin{equation}
|\Delta^+ \uparrow\ra=\frac{1}{3}(uud+udu+duu)(\uparrow\uparrow\downarrow
+\uparrow\downarrow\uparrow+\downarrow\uparrow\uparrow)
\end{equation}
Now we substitute Eq. (\ref{field}) into Eq. (\ref{hint}) and the latter into 
Eq. (\ref{amp}). Then, using the states given above, 
we calculate the sandwiches of $H$, obtaining the final amplitude. 
The cross section for the process $p \to \Delta$ reads:
\begin{equation}
    \sigma=\int |a_{fi}|^2 \, d^2b 
=\frac{Z^2e^4}{9\pi m^2}
  \left( \frac{E_{fi}}{v\gamma} \right)^2 \int_{R}^{\infty}
\Big[ K_1\Big(\frac{E_{fi}b}{v\gamma}\Big)\Big]^2 b \, db
\label{sigclass}
\end{equation}
For our purpose it is sufficient to have only one proton as target.

In the formalism developed in \cite{nos21} the cross section of reaction 
depicted in Fig.~\ref{fig3}b  reads \cite{upc}:
\begin{equation}
\sigma = \int \frac{d\omega}{\omega} \, n(\omega)
\,  \sigma_{\gamma N\rightarrow N\pi} (\omega)
\label{sigquan}
\end{equation} 
where $n(\omega)$ is given by \cite{upc}:
\begin{equation}
  n(\omega) = \frac{Z^2 \alpha}{\pi}\Bigg[2\xi K_0(\xi)K_1(\xi)-\xi^2[K_1^2(\xi)
      -K_0^2(\xi)]\Bigg],  \hspace{2cm} \xi=\frac{\omega (R_1+R_2)}{\gamma}
\label{flux}
\end{equation}
In the above expression  $\omega$ is the photon energy, $R_i$ is the radius of 
nucleus $i$. The Lorentz $\gamma$  factor is  
in the target frame. In the LHC  $\gamma \simeq 1000$.

\begin{figure}[h]                                                              
\centering                                                                     
\includegraphics[width=.50\linewidth]{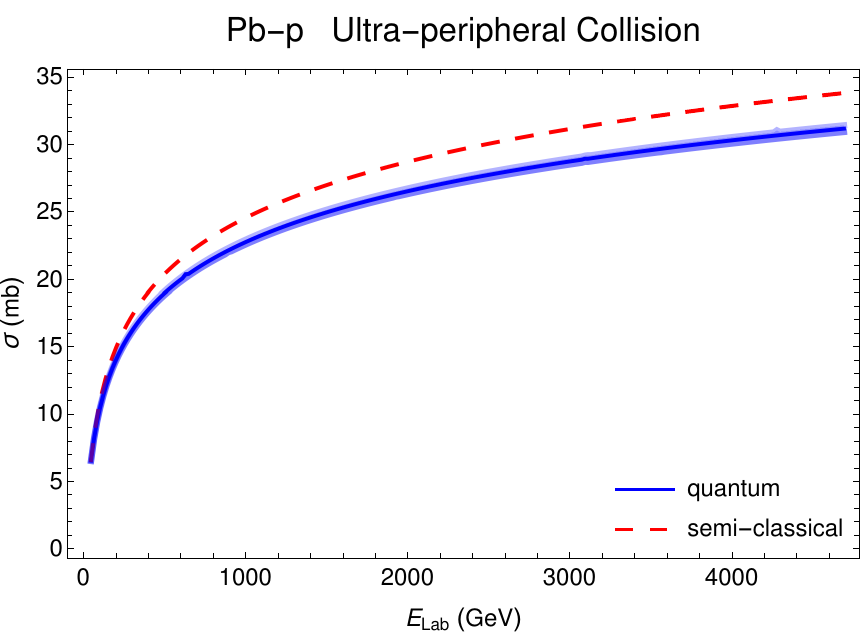}                            
\caption{Cross sections for pion production obtained with the semi-classical 
formalism, Eq.(\ref{sigclass}), dashed line, and with the quantum formalism, 
Eq. (\ref{sigquan}), solid line. $E_{Lab} = \gamma \, m_n$ is  
the energy per nucleon  in the laboratory frame. }
\label{fig4}                                                                    
\end{figure}

The cross section of the process $\gamma p \rightarrow p \pi$                
can be computed from the Feynman graph  depicted in Fig. \ref{fig3}b.  
A simple parametrization of the $\pi^0$  photoproduction 
cross section was introduced in  \cite{nos21}. 
Knowing  $\sigma_{\gamma N \rightarrow N \pi}$,  we use it to calculate the
cross section (\ref{sigquan}).  In Fig. \ref{fig4} we plot together the 
obtained quantum (\ref{sigquan}) (solid lines) and semi-classical 
(\ref{sigclass}) (dashed line) cross sections.  
The band  shows the uncertainty related to the decay 
width $\Gamma$ \cite{nos21}. The difference between the results obtained with 
(\ref{sigclass}) and with (\ref{sigquan}) is small and reaches 9 \% 
at the highest energies. 

The measurement of these ultra-forward pions may be challenging, but there 
is some hope. In fact, ultra-forward neutral pions have already been
detected in proton-proton and proton-lead collisions at the LHC \cite{lhcf}.  
Unfortunately, in those measurements it was not possible to focus only on 
ultra-peripheral collisions. We hope that this  could be done in the future. 

\section{Conclusion}

In this work  the cross section for the production of a $D^+ D^-$ 
molecule in ultra-peripheral collisions was calculated.  
It is $\sigma_B(AA \to AAB) = 3.0^{+0.8}_{-1.2} \,\, \mu b $  
for $\sqrt{s_{NN}} = 5.02$ TeV. This number is consistent with the  results
obtained for other scalar exotic charmonium molecules in Ref.~\cite{br} and
in Ref.~\cite{fa}. The  parameters of the calculation  are the hadronic 
form factor 
cut-off, the maximum transverse momentum of an emitted photon and the 
binding energy.  
All these parameters can be constrained by experimental information and/or 
by calculations and hence the precision of our calculation can be increased.
The method used here can be easily applied to other exotic states. 

We have also calculated the cross section for the production of very forward
pions. We have used two methods, one with a classical magnetic field and the
other with equivalent photons. Both methods yield a similar result: a quite 
large cross section for forward pion production. The neutral pions can in 
principle be measured. This would improve our knowledge about the validity of
the classical approximation and about the strength of the magnetic field 
created in these collisions.

\section*{Acknowledgments}
We are deeply indebted to K. Khemchandani, A. Martinez Torres, A. Szczurek and
A. Esposito for instructive discussions.
This work was  partially financed by the Brazilian funding
agencies CNPq, CAPES, FAPESP,  FAPERGS and INCT-FNA (process number
464898/2014-5). F.S.N.  gratefully acknowledges the  support from the
Funda\c{c}\~ao de Amparo \`a  Pesquisa do Estado de S\~ao Paulo (FAPESP).

%


\end{document}